\documentclass[sigconf,natbib=true]{acmart}
\AtBeginDocument{%
  \providecommand\BibTeX{{%
    Bib\TeX}}}

\copyrightyear{2026}
\acmYear{2026}
\setcopyright{cc}
\setcctype{by}
\acmConference[SIGIR '26]{Proceedings of the 49th International ACM SIGIR Conference on Research and Development in Information Retrieval}{July 20--24, 2026}{Melbourne, VIC, Australia}
\acmBooktitle{Proceedings of the 49th International ACM SIGIR Conference on Research and Development in Information Retrieval (SIGIR '26), July 20--24, 2026, Melbourne, VIC, Australia}
\acmDOI{10.1145/3805712.3808587}
\acmISBN{979-8-4007-2599-9/2026/07}

\usepackage[utf8]{inputenc} 
\usepackage[T1]{fontenc}    
\usepackage{hyperref}       
\usepackage{url}            
\usepackage{booktabs}       
\usepackage{amsfonts}       
\usepackage{nicefrac}       
\usepackage{microtype}      
\usepackage{xcolor}         
\usepackage{graphicx}
\usepackage{subfigure}
\usepackage{cleveref}
\usepackage{caption}
\usepackage{threeparttable}
\usepackage{comment}
\usepackage{multirow}
\usepackage{enumitem}
\usepackage{adjustbox}
\usepackage{bbding}
\usepackage[normalem]{ulem}
\useunder{\uline}{\ul}{}
\usepackage{latexsym}
\usepackage{algorithm}
\usepackage{algorithmic}
\usepackage{mathrsfs}
\usepackage{balance}
\usepackage{dcolumn}
\usepackage{tabularx}
\usepackage{colortbl}
\usepackage{xspace,mfirstuc,tabulary}
\usepackage{amsthm}
\usepackage{ulem}
\usepackage{multirow,booktabs, hhline}
\usepackage{bm}
\usepackage{color}
\usepackage{pifont}

\newenvironment{itemize*}%
 {\leftmargini=20pt\begin{itemize}%
  \setlength{\itemsep}{3pt}%
  \setlength{\parskip}{0pt}%
  }%
 {\end{itemize}}
\newenvironment{enumerate*}%
 {\begin{enumerate}%
  \setlength{\itemsep}{0pt}%
  \setlength{\parskip}{0pt}}%
 {\end{enumerate}}
\usepackage{amsmath}
\usepackage{makecell}
\usepackage{bbm}
\usepackage{CJKutf8}
\usepackage{listings}
\usepackage{wrapfig}
\usepackage{lipsum}
\usepackage{pgfplots}
\usepackage{tikz}
\usetikzlibrary{calc}
\usepackage{array}
\usepackage{arydshln}
\usepackage{multicol}

\definecolor{chart Idle}{gray}{.6}
\definecolor{chart Poor}{RGB}{242,28,28}
\definecolor{chart Ok}{RGB}{248,172,37}
\definecolor{chart Ideal}{RGB}{1,151,0}
\definecolor{chart Over}{RGB}{0,125,234}
\definecolor{lightergray}{RGB}{230,230,230}
\definecolor{DarkGreen}{RGB}{30,130,30}

\newcommand\ourdata{KuaiLive\xspace}

\newcommand{\paratitle}[1]{\vspace{1.5ex}\noindent\textbf{#1}}

\newcommand{\eg}{\emph{e.g.,}\xspace}

\newcommand{\etc}{\emph{etc}}

\newdimen\tempdim

\newcommand*{\ChartBox}[3]{%
  \begingroup
    \settoheight{\tempdim}{L}%
    \edef\tempheight{\the\tempdim}%
    \settodepth{\tempdim}{g}%
    \edef\tempdepth{\the\tempdim}%
    \tikz[
      baseline=0pt,
      inner sep=0pt,
    ]
    \node[
      fill={#3!#2},
      rounded corners=1pt,
      anchor=base,
    ]{%
      \vphantom{g\"A}%
      \pgfmathsetlength{\tempdim}{#1}%
      \kern\tempdim\relax
    };%
  \endgroup
}

\AtBeginDocument{%
  \providecommand\BibTeX{{%
    \normalfont B\kern-0.5em{\scshape i\kern-0.25em b}\kern-0.8em\TeX}}}
\begin{document}
\begin{sloppy}

\title{‌KuaiLive: A Real-time Interactive Dataset for \\Live Streaming Recommendation‌}

\author{Changle Qu}
\author{Sunhao Dai}
\affiliation{
  \institution{\mbox{Gaoling School of Artificial Intelligence}\\Renmin University of China}
  \city{Beijing}
  \country{China}
  }
\email{{changlequ, sunhaodai}@ruc.edu.cn}

\author{Ke Guo}
\author{Xiao Zhang}
\authornote{Corresponding author: Xiao Zhang (e-mail: zhangx89@ruc.edu.cn). }
\affiliation{
  \institution{\mbox{Gaoling School of Artificial Intelligence}\\Renmin University of China}
  \city{Beijing}
  \country{China}
  }
\email{{guoke0228, zhangx89}@ruc.edu.cn}

\author{Liqin Zhao}
\author{Shijun Wang}
\affiliation{
  \institution{Kuaishou Technology}
  \city{Beijing}
  \country{China}
  }
\email{{zhaoliqin,wangshijun03}@kuaishou.com}

\author{Yanan Niu}
\author{Lantao Hu}
\affiliation{
  \institution{Kuaishou Technology}
  \city{Beijing}
  \country{China}
  }
\email{{niuyanan,hulantao}@kuaishou.com}

\author{Han Li}
\affiliation{
  \institution{Kuaishou Technology}
  \city{Beijing}
  \country{China}
  }
\email{lihan08@kuaishou.com}

\author{Jun Xu}
\affiliation{
  \institution{\mbox{Gaoling School of Artificial Intelligence}\\Renmin University of China}
  \city{Beijing}
  \country{China}
  }
\email{junxu@ruc.edu.cn}

\renewcommand{\shorttitle}{KuaiLive: A Real-time Interactive Dataset for Live Streaming Recommendation}
\renewcommand{\shortauthors}{Changle Qu et al.}
\begin{abstract}
Live streaming platforms have become a dominant form of online content consumption, offering dynamically evolving content, real-time interactions, and highly engaging user experiences. 
These unique characteristics introduce new challenges that differentiate live streaming recommendation from traditional recommendation settings and have garnered increasing attention from industry in recent years.
However, research progress in academia has been hindered by the lack of publicly available datasets that accurately reflect the dynamic nature of live streaming environments.
To address this gap, we introduce \ourdata, the first real-time, interactive dataset collected from Kuaishou, a leading live streaming platform in China with over 400 million daily active users.
The dataset records the interaction logs of 23,772 users and 452,621 streamers over a 21-day period. 
Compared to existing datasets, \ourdata offers several advantages: it includes precise live room start and end timestamps, multiple types of real-time user interactions (click, comment, like, gift), and rich side information features for both users and streamers. 
These features enable more realistic simulation of dynamic candidate items and better modeling of user and streamer behaviors.
We conduct a thorough analysis of \ourdata from multiple perspectives and evaluate several representative recommendation methods on it, establishing a strong benchmark for future research.
\ourdata can support a wide range of tasks in the live streaming domain, such as top-$K$ recommendation, click-through rate prediction, watch time prediction, and gift price prediction. 
Moreover, its fine-grained behavioral data also enables research on multi-behavior modeling, multi-task learning, and fairness-aware recommendation.
We believe that \ourdata will serve as a valuable resource to advance the development of intelligent live streaming services. 
The dataset and related resources are publicly available at~\textcolor{blue}{\url{https://imgkkk574.github.io/KuaiLive}}.
\end{abstract}



\maketitle
 
\section{Introduction}
\label{introsection}
\begin{figure}[t]  
    \centering    
    \includegraphics[width=1\columnwidth]{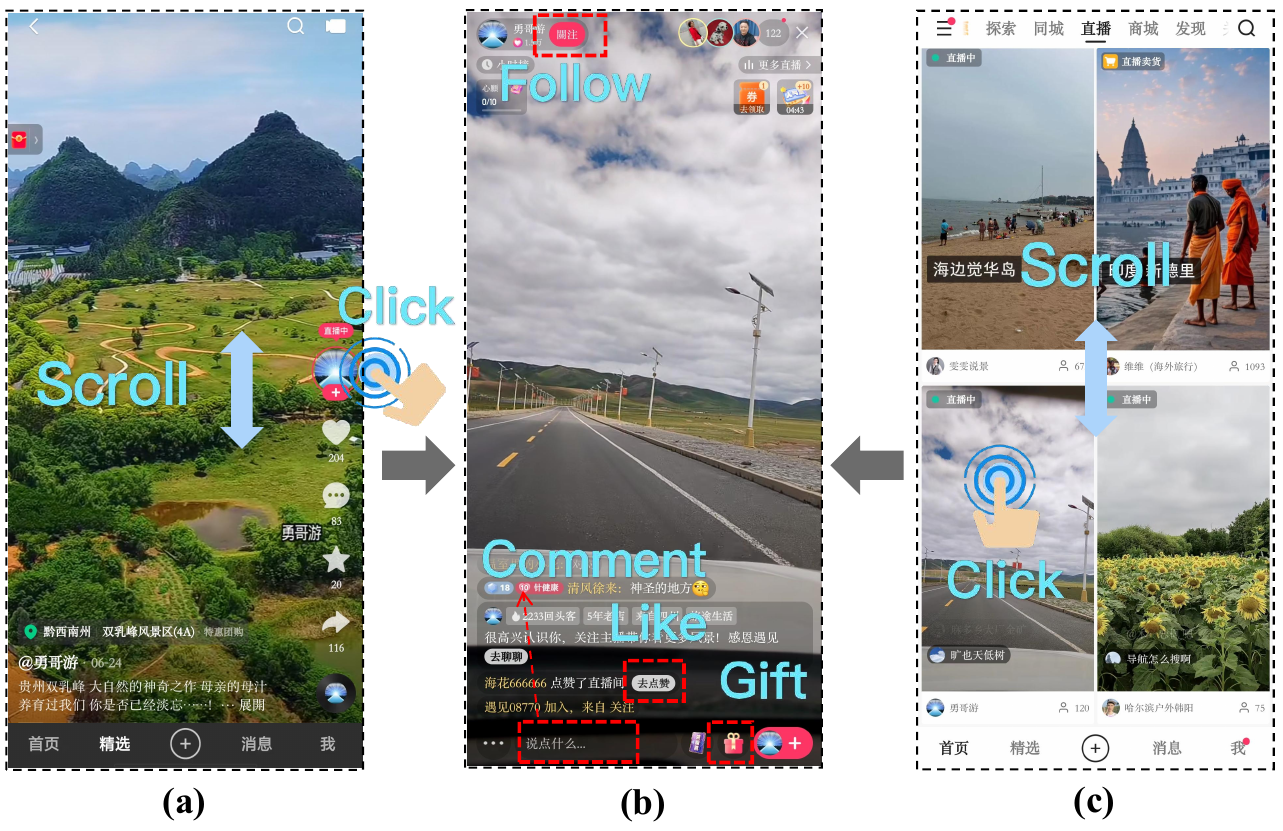}
    \caption{Illustration of live streaming scenarios in Kuaishou App. (a)~The single-column recommendation feed, where users scroll vertically to receive a mix of short videos and live streams. (b) The live streaming interface, where users can interact with the streamer through actions such as Follow, Comment, Like, and Gift. (c) The two-column live streaming recommendation interface, where users scroll to browse live streams and click a thumbnail to enter a live room.}
    \label{fig:analy_sim} 
\end{figure}

In the era of rapidly evolving streaming services, streaming recommendation platforms such as Kuaishou have not only attracted widespread attention, but also become core application scenarios for streaming machine learning~\cite{yu2021leveraging,dai2024recode,zhang2021counterfactual,zhang2022counteracting,guo2026room,shen2026enhancing,zhang2023reward}. 
Among these, live streaming has emerged as a novel form of online service, seamlessly integrating real-time content broadcasting with immediate social engagement~\cite{liang2024ensure,deng2024mmbee,qu2025bridging}. 
In a typical live room, streamers share content in real time while users actively express their preferences and support through comments, likes, or virtual gifts, fostering a highly dynamic and participatory environment.
Despite its rising popularity and practical importance, live streaming recommendation remains significantly underexplored in academic research, primarily due to the lack of publicly available, large-scale, and well-structured datasets that support systematic investigation and benchmarking.
Consequently, a substantial research gap has emerged between industry practice and academic understanding, hindering the development of effective and generalizable recommendation models tailored to the unique characteristics of live streaming scenarios.

Existing representative public recommendation datasets such as KuaiRec~\cite{gao2022kuairec}, KuaiSAR~\cite{sun2023kuaisar}, and Tenrec~\cite{yuan2022tenrec} primarily focus on short video or general content recommendation tasks. 
However, live streaming recommendation differs significantly from traditional recommendation scenarios, exhibiting several unique characteristics:
(1)~\textbf{Streamers and live rooms are only active during broadcast periods}, resulting in a time-dependent and dynamic candidate pool that changes continuously over time.
(2)~\textbf{Both live content and user behavior evolve in real time}, users respond to varying content with diverse feedback signals, such as comments, likes, and virtual gifts, leading to complex and temporally correlated interaction patterns.
Even datasets specifically designed for live streaming recommendation such as LiveRec~\cite{rappaz2021recommendation}, LSEC~\cite{yu2021leveraging}, and KLive~\cite{deng2024multimodal} fail to comprehensively capture these dynamic properties. 
This highlights the urgent need for a more representative dataset that reflects the true complexity of live streaming interactions.

To bridge this gap, we introduce \ourdata, a large-scale real-world dataset for live streaming recommendation, collected from Kuaishou\footnote{https://www.kuaishou.com}.
On Kuaishou, users can discover and enter live rooms of interest, where they interact with streamers in real time through behaviors such as clicking, liking, commenting, following, and sending virtual gifts (as shown in Figure~\ref{introsection}). 
These rich user–streamer interactions make Kuaishou an ideal source for constructing a realistic live streaming recommendation dataset.
Compared with existing public datasets, \ourdata offers several notable advantages:
(1) It includes the start and end timestamps of each live room, allowing researchers to simulate realistic live streaming recommendation settings where candidate items are temporally constrained and dynamically changing.
(2) It records multiple user behaviors (e.g., click, comment, like, gift), which can be leveraged to study multi-task learning and multi-behavior modeling.
(3) It preserves the temporal order of each interaction, supporting fine-grained analysis of user behavior trajectories.
(4) It includes user watch time and gift price, enabling broader research tasks beyond recommendation, such as watch time and gift price prediction.
(5) It contains not only positive feedback but also negative feedback, making it suitable for click-through rate (CTR) prediction.
(6) It provides not only user and item IDs but also rich side information features.

Notably, \ourdata is the first publicly available live streaming dataset that captures rich and realistic sequences of user interactions within an interactive environment. This makes it a valuable resource for advancing research in live streaming recommendation, including but not limited to top-$K$ recommendation and CTR prediction.
In addition, thanks to its fine-grained behavioral information, \ourdata also enables a wide range of research tasks, such as watch time prediction, gift price prediction, and multi-task learning, \etc.

\begin{table}[t]
\centering
\caption{Comparison between \ourdata and existing public live streaming recommendation datasets. Symbols \ding{172}, \ding{173}, \ding{174}, \ding{175}, and \ding{176} represent the five types of action in live room—click, follow, comment, like, and gift, respectively.}
\label{tab:comparison_statistics}
\begin{threeparttable}
\begin{small}
\begin{tabular}{l|cccc}
\Xhline{0.7pt}
\textbf{Property} & \textbf{LiveRec}$^{1}$ & \textbf{LSEC}$^{2}$ & \textbf{Klive} & \textbf{\ourdata}\\
\hline
\# Users & 15,500,000 & 202,850 & - & 23,772\\
\# Streamers & 465,000 & 7,395 & 9,932 & 452,621 \\ 
\# Rooms & ? & - & 17,798 & 11,613,708 \\ 
\# Interactions & 124,000,000 & 5,439,288 & - & 5,357,998\\
\# User features &0 & 0 & 0 & 20\\
\# Streamer features &0 & 0 & 0 & 23\\
Lifecycle &$\times$ & $\times$ & $\times$ & $\surd$\\
Negative &$\times$ & $\times$ & $\times$ & $\surd$\\
Action type &\ding{172} & \ding{173} & - & \ding{172}, \ding{174}, \ding{175}, \ding{176}\\
\Xhline{0.7pt}
\end{tabular}
\end{small}
\begin{tablenotes}
\footnotesize
	\item[1] LiveRec has two versions: Bench. and Full. The reported statistics are based on the Full version, which includes live room information, although such details are not explicitly presented in the paper and require manual extraction.
	\item[2] LSEC has two versions: Small and Large. The statistics presented here are based on the Large version, which includes only user-streamer interactions.
\end{tablenotes}
\end{threeparttable}
\end{table}

\section{Related Work}
\paratitle{Live Streaming Recommendation.} 
Recently, live streaming has become an important form of social media and has attracted growing research interest~\cite{yu2021leveraging,zhang2023cross,rappaz2021recommendation,zhu2025live,lu2025liveforesighter}.
Given the multi-modal nature of live streaming, early works aim to integrate textual, visual, and acoustic signals to better model users and live content~\cite{xi2023multimodal,deng2023contentctr,deng2024mmbee,liu2025llm,guo2026room}.
Other studies investigate real-time content evolution and highlight detection within live rooms~\cite{liang2024ensure,deng2024multimodal}.
Considering that many platforms provide both short video and live streaming services, recent efforts explore cross-domain preference transfer between short videos and live streams to improve recommendation quality~\cite{cao2024moment,li2025farm,qu2025bridging}.
Despite these advancements, a major bottleneck persists: most existing methods are evaluated on proprietary industrial datasets. This lack of public benchmarks limits reproducibility and hinders academic progress, highlighting the urgent need for the open-source dataset presented in this work.

\begin{table*}[t]
 \caption {Statistics of our proposed \ourdata (top) and feature descriptions (bottom).
}
\label{tab: statistics} 
\tabcolsep=7.5pt 
\begin{tabular}{l|cccccccc}
\Xhline{0.7pt}
 Dataset & \#Users & \#Streamers & \#Rooms &  \#Interactions  &\#Clicks  &\#Comments &\#Likes &\#Gifts \\ \hline
 \ourdata & 23,772  & 452,621 & 11,613,708   & 5,357,998  &4,909,515 &196,526 &179,311 &72,646\\
\Xhline{0.7pt}
\end{tabular}

\vspace{5pt}

\tabcolsep=7.5pt
\setlength{\tabcolsep}{4pt}
\begin{tabular}{@{}ll@{}}
\Xhline{0.7pt}
\textbf{User}:  & \begin{tabular}[c]{@{}l@{}} gender, age, country, device\_brand, device\_price, reg\_timestamp, fans\_num, follow\_num, is\_video\_author,\\first\_watch\_live\_timestamp, accu\_watch\_live\_cnt, accu\_watch\_live\_duration, is\_live\_author, 7 encrypted vectors. \end{tabular}                                                                                   \\ \midrule
\textbf{Streamer}: & \begin{tabular}[c]{@{}l@{}} gender, age, country, device\_brand, device\_price, reg\_timestamp, live\_operation\_tag, first\_live\_timestamp, fans\_group\_num, \\fans\_num, follow\_num, accu\_live\_cnt, accu\_live\_duration, accu\_play\_cnt, accu\_play\_duration, 7 encrypted vectors. \end{tabular}                                                                                   \\ \midrule
\textbf{Room}:  & \begin{tabular}[c]{@{}l@{}} start\_timestamp, end\_timestamp, live\_type, live\_content\_category, live\_name\_representation. \end{tabular}                                                                                   \\ \Xhline{0.7pt}
\end{tabular}
\end{table*}

\paratitle{Related Datasets.} 
Datasets serve as the foundation for both developing and evaluating recommendation algorithms. 
To the best of our knowledge, only three public datasets are available for live streaming recommendation:
LiveRec~\cite{rappaz2021recommendation}, which records user watch time; 
LSEC~\cite{yu2021leveraging}, an e-commerce live streaming dataset capturing behaviors such as following and purchasing; 
and Klive~\cite{deng2024multimodal}, which includes textual and multimodal features of live sessions.
Despite their contributions, these datasets have notable limitations: they lack live room lifecycle information, fail to capture diverse interaction types, and are often restricted to ID-level fields.
As a result, they fail to fully capture the dynamic and interactive nature of real-world live streaming scenarios.
The comparison between \ourdata and these datasets is presented in Table~\ref{tab:comparison_statistics}.
Compared with existing datasets, \ourdata offers a more comprehensive and fine-grained view of the live streaming ecosystem by simultaneously incorporating information from three parties: users, streamers, and rooms, along with multiple types of user interactions.
It also records the lifecycle of each live room,  allowing for modeling of the evolving candidate set in real scenarios.
Furthermore, it includes rich side information and diverse interaction types, covering both positive (\eg click, comment, like, gift) and negative feedback.

\section{The KuaiLive Dataset}
In this section, we present a comprehensive overview of the KuaiLive dataset. 
We begin by introducing the characteristics of the Kuaishou app to provide background and context for the dataset. 
Next, we describe the data construction process in detail. 
Finally, we summarize the KuaiLive dataset by highlighting its scale and the diverse features it contains. 

\subsection{Characteristics of Kuaishou App}
Kuaishou is one of the largest short video and live streaming platforms in China, boasting over 400 million daily active users.
In recent years, live streaming has become an integral part of daily life and is emerging as one of Kuaishou’s core services. 
Notably, revenue from live streaming contributes 30\% of the company’s total income, highlighting its significant commercial potential.

Unlike platforms such as Taobao\footnote{https://tbzb.taobao.com/}, which focuses on e-commerce live streaming, or Twitch, which is centered on gaming content, Kuaishou offers a diverse range of live streaming content spanning e-commerce, education, gaming, fitness, and more. 
This diversity has attracted a broad user base and fostered active engagement in live interactions. 
Such a solid ecosystem provides a strong foundation for constructing the live streaming recommendation dataset.

We illustrate the live streaming service on Kuaishou in Figure~\ref{introsection}. 
On the right side, the live streaming section adopts a two-column layout, where users can scroll vertically to explore recommended live streams and click a thumbnail to enter a live room of interest. 
On the left side, Kuaishou integrates live streams with short videos in a unified single-column feed, where each item can be either a short video or a live stream.
While browsing short videos, users may enter a live room by clicking the creator’s profile if the streamer is currently live.
Once in a live stream, users can interact with the streamer through actions such as following, liking, commenting, and sending virtual gifts, all of which are recorded in \ourdata. 

\begin{figure*}[t]
    \centering
    \subfigtopskip=2pt 
	\subfigbottomskip=2pt 
	\subfigcapskip=-5pt 
    \subfigure[User Engagement Distribution.]{
    \includegraphics[width=0.266\linewidth]{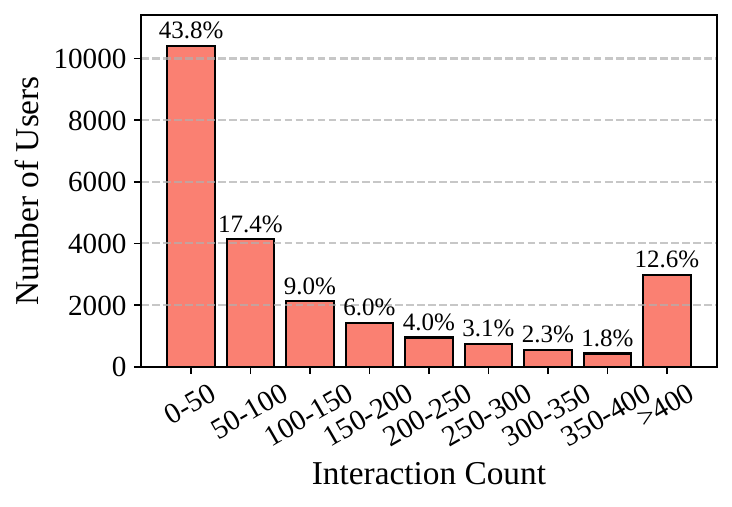}
    }
    \subfigure[Temporal User Activity.]{
    \includegraphics[width=0.427\linewidth]{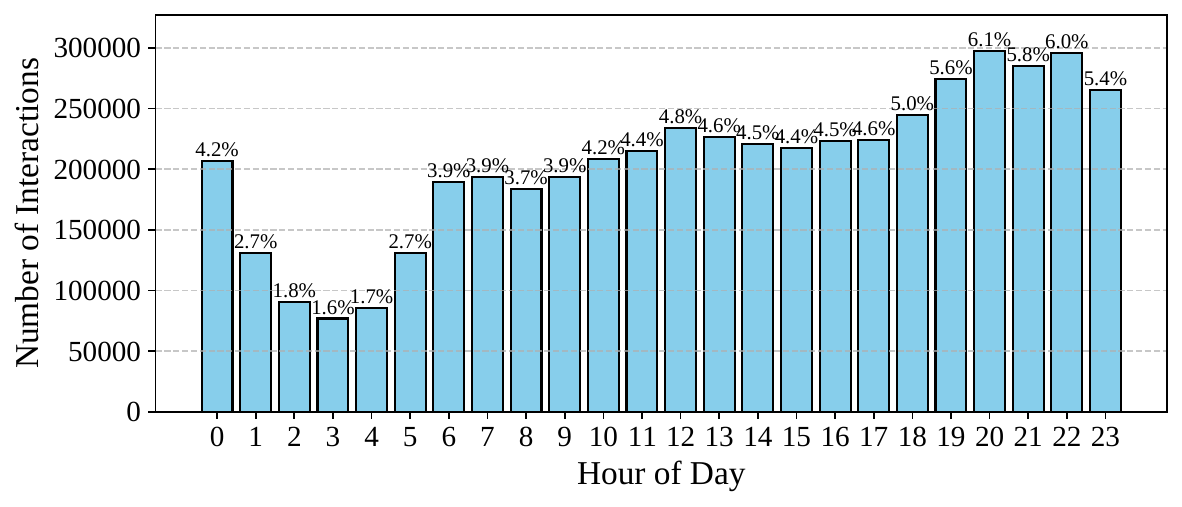}
    }
    \subfigure[Repeat Consumption Behavior.]{
    \includegraphics[width=0.266\linewidth]{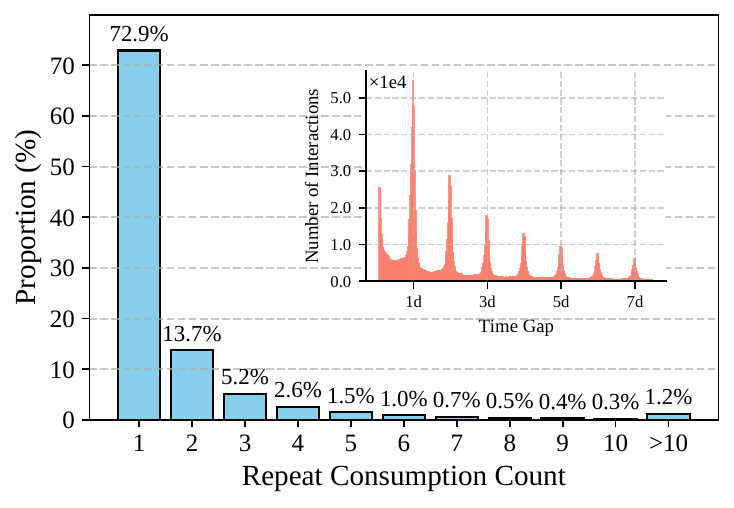}
    }
    \caption{Analysis of user interactions in the \ourdata dataset. (a) shows the distribution of interaction counts per user. (b) illustrates the hourly distribution of user interactions over a 24-hour period. (c) shows the distribution of user interactions with the same streamer and the time gap between repeat consumption, revealing a clear daily periodic pattern.}
    \label{fig:analy_user}
\end{figure*}

\subsection{Data Construction}
To facilitate the research on live streaming recommendation, \ourdata is constructed with the following steps:

\textbf{User Sampling.}
To ensure both the representativeness and behavioral diversity, we randomly sample approximately 25,000 active users who engaged in all four types of interactions~(click, comment, like, and gift) in the live streaming domain of the Kuaishou between May 5, 2025, and May 25, 2025. 
After filtering out users with abnormal behaviors, we retain a total of 23,772 unique user IDs.

\textbf{Interaction Collection.}
We collect fine-grained user interactions from live streaming logs, including four behavior types: click, comment, like, and gift, each with precise timestamps.
To avoid interest bias from pre-existing follows, follow behavior is excluded from this dataset.
In addition to explicit interactions, we record auxiliary signals (e.g., watch time and gift price) and negative feedback from exposed-but-skipped rooms, resulting in interactions involving 452,621 streamers and 11,613,708 live rooms.

\textbf{Side Information Collection.}
To further enhance the utility and realism of~\ourdata, we collect extensive side information for all three core entities: users, streamers, and live rooms.
As summarized in Table~\ref{tab: statistics}, the dataset includes 20 user features, 23 streamer features, and 5 room features, covering a diverse set of attributes such as demographics, content characteristics, and behavioral summaries.
For time-sensitive features such as fans\_num, we standardize their values by taking consistent snapshots as of May 25, 2025, to ensure temporal coherence across the dataset.
With these features, researchers can explore a broader range of tasks using this dataset. 

\textbf{Anonymization.}
Given that \ourdata is collected from a commercial platform, strict anonymization is applied to ensure compliance with data-releasing policies and protect user privacy. 
Specifically, the IDs of users, streamers, and rooms are randomly hashed into anonymized integer values, thereby removing any direct traceability.
In addition, textual data such as live room titles are first encoded into dense vectors using a pre-trained embedding model, and then reduced in dimensionality to prevent potential information leakage. These anonymization strategies effectively safeguard the dataset from containing any personally identifiable or sensitive content, while preserving the data's utility for research purposes.

\subsection{Statistics}
We summarize the key statistics of \ourdata in Table~\ref{tab: statistics}.
The dataset consists of 23,772 users, 452,621 streamers, and 11,613,708 live rooms, noting that while each streamer has at least one user interaction, only a subset of their live rooms receive engagement.
Over the 21-day collection period, \ourdata records 5,357,998 user interactions across four behavior types: click, comment, like, and gift.
Compared with clicks, other behaviors are much rarer, especially gifts, which account for only 1.5\% of clicks, reflecting the high cost of virtual gifting and the resulting extreme sparsity.
Such a distribution highlights the extreme sparsity of gifting interactions in live streaming scenarios, reflecting the inherent challenges of modeling user intent and engagement in such scenarios.
We believe this underexplored yet practically important scenario deserves more attention from the research community. 

\subsection{\ourdata License}
The dataset is available for non-commercial use under a custom Creative Commons Attribution-NonCommercial-ShareAlike 4.0 International License\footnote{https://creativecommons.org/licenses/by-nc-sa/4.0/} (CC BY-NC-SA 4.0).

\begin{figure*}[t]
    \centering
    \subfigtopskip=2pt 
	\subfigbottomskip=2pt 
	\subfigcapskip=-5pt 
    \subfigure[Streamer Engagement Distribution.]{
    \includegraphics[width=0.266\linewidth]{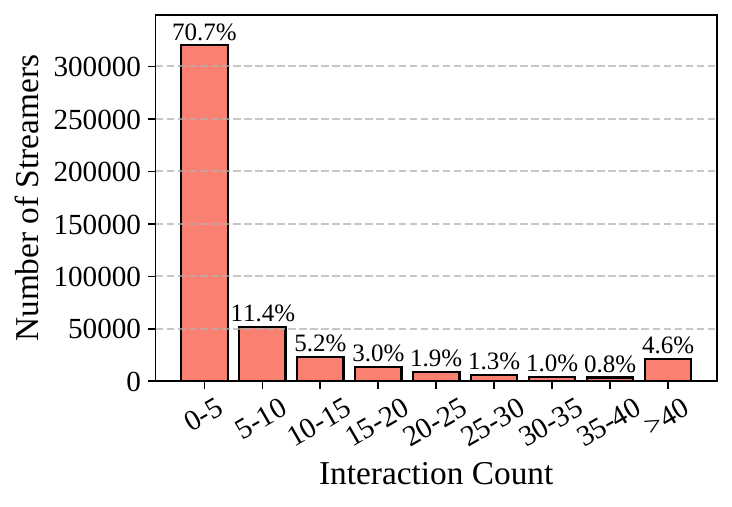}
    }
    \subfigure[Long-tail Popularity Distribution.]{
    \includegraphics[width=0.266\linewidth]{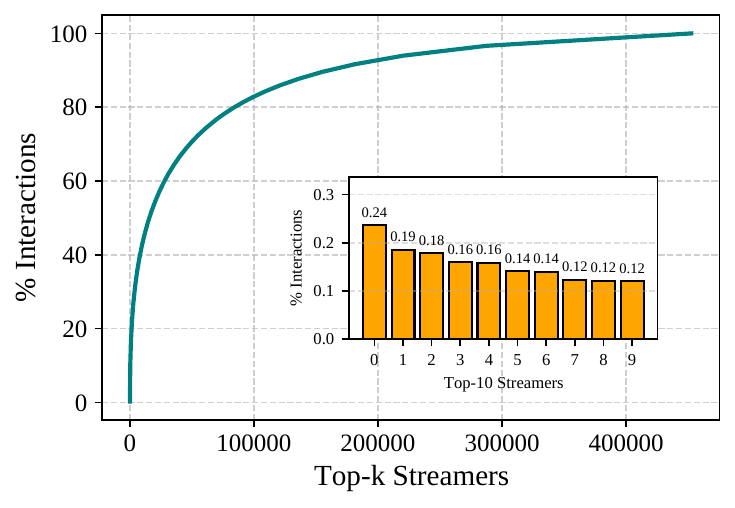}
    }
    \subfigure[Temporal Streaming Behaviors.]{
    \includegraphics[width=0.427\linewidth]{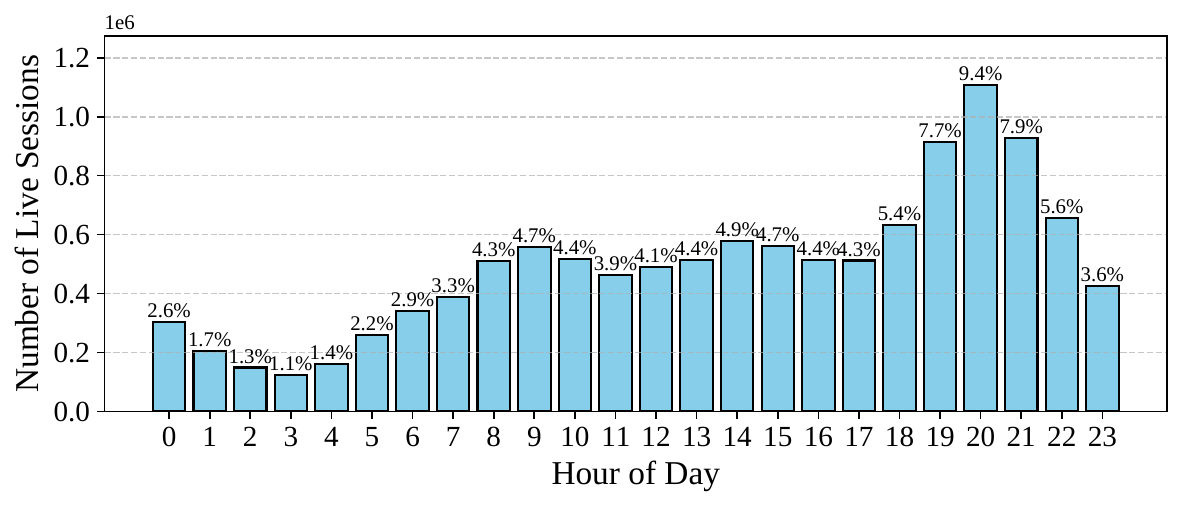}
    }
    \caption{Analysis of streamer behaviors in the \ourdata dataset. (a) shows the distribution of interaction counts per streamer. (b) illustrates the cumulative distribution of streamer interactions, showing that the top-10 most popular streamers account for approximately 1.5\% of total interactions. (c) shows the hourly distribution of streamer start timestamps over a 24-hour period.
}
    \label{fig:analy_streamer}
\end{figure*}

\section{Dataset Analysis}
In this section, we conduct a comprehensive analysis of the \ourdata dataset from three aspects: 1) demographics, 2) streamer activity analysis, and 3) user interaction behaviors. 
This analysis helps reveal key characteristics of the live streaming scenario and provides valuable insights for downstream recommendation tasks.

\subsection{Demographics}
We first analyze the demographics of the 23,772 users and 452,621 streamers in \ourdata.
Among users, 61.85\% are male, while the remaining 38.15\% are female, slightly more male-skewed compared to the overall user base of the app. 
In contrast, streamers exhibit an opposite trend, with 62.83\% female and 37.17\% male.
Additionally, \ourdata covers a wide variety of 13 streamer types, including Chat, E-Commerce, Beauty, Lifestyle, Talent, Education, Relationship, Game, Hobbies, Fitness, Group, News, and Other.
Since the data is collected from the Chinese version of the Kuaishou app rather than the international version (Kwai), the majority of users and streamers are based in China. 
While the dataset includes IP addresses from 83 different countries, over 99\% of users and streamers are from China, with a small portion coming from South Korea and Japan.

\subsection{Streamer Activity Analysis}
\label{streamer_analy}
In this section, we investigate the behavioral characteristics and popularity patterns of streamers in the \ourdata dataset:

\textbf{Streamer Engagement Distribution.}
First, we analyze the distribution of interaction frequency per streamer in the \ourdata dataset, as shown in Figure~\ref{fig:analy_streamer}(a). 
The results exhibit a severe imbalance: approximately 70\% of streamers receive fewer than 5 interactions, while only 4.6\% receive more than 40.
This indicates a prevalent cold-start problem for most streamers, posing challenges for visibility and engagement.
Moreover, it suggests that commonly adopted preprocessing strategies such as 5-core filtering method may distort real-world behavior and lead to discrepancies between models trained on filtered and full datasets.

\textbf{Long-tail Popularity Distribution.}
Building on the above observation, we examine the cumulative distribution of interactions per streamer in Figure~\ref{fig:analy_streamer}(b), confirming a pronounced long-tail effect.
The results reveal a highly skewed distribution, where a small fraction of top streamers dominate user attention. 
Notably, the top 10 most popular streamers alone account for over 1.5\% of all interactions in the \ourdata. 
Such head-heavy dynamics pose challenges for fairness-aware recommendation, as new or less-followed streamers may struggle to gain exposure.
Addressing this imbalance is crucial to avoid discouraging emerging creators and to support a more diverse and sustainable live streaming ecosystem.

\textbf{Temporal Streaming Behaviors.}
Finally, we analyze streamers’ activity across different times of the day by examining the hourly distribution of live stream initiations.
As shown in Figure~\ref{fig:analy_streamer}(c), the number of live rooms rises sharply in the evening, peaking between 6 PM and 10 PM, while far fewer streams are initiated between 11 PM and 8 AM.
This pattern suggests that streamers schedule broadcasts during periods of high user availability to maximize engagement.
Such temporal variation results in uneven candidate distributions across time slots, highlighting the need for recommender systems to explicitly incorporate temporal dynamics.

\subsection{User Interaction Analysis}
In this section, we analyze the behavioral characteristics of users:

\textbf{User Engagement Distribution.}
We first examine the overall distribution of user interactions. 
As shown in Figure~\ref{fig:analy_user}(a), although the imbalance is less severe compared to streamers, it still remains prominent: 43.8\% of users have fewer than 50 interactions, while 12.6\% of users are highly active. 
This presents two main challenges for recommender systems. 
For users with limited interactions, it is difficult to construct accurate user profiles due to the cold-start problem. 
On the other hand, for highly active users with diverse behaviors, identifying their core interests from a large and potentially noisy interaction history becomes equally challenging.

\textbf{Temporal User Activity.}
Next, we analyze user engagement over a 24-hour period to capture temporal variations in activity.
As shown in Figure~\ref{fig:analy_user}(b), user interactions peak in the evening, particularly between 6 PM and 12 AM, closely aligning with streamers’ active hours, while engagement remains low in the early morning.
Interestingly, there is also a smaller peak around 12 PM, which may correspond to users watching live streams during their lunch break as a form of relaxation.
These observations suggest that user behavior is closely tied to daily routines and that user preferences may vary across different times of the day. 
For instance, food-related streams may gain more traction around noon, while gaming content often dominates in the evening, when users are more inclined to engage in leisure-oriented activities after work or school.
Such temporal dynamics highlight the importance of incorporating time-sensitive user preferences into real-time recommender systems.

\textbf{Repeat Consumption Behavior.}
Finally, we examine repeat consumption behaviors in the live streaming scenario to gain deeper insights into user engagement patterns.
As illustrated in Figure~\ref{fig:analy_user}(c), 27.1\% of users interact with the same streamers multiple times, indicating that repeat consumption is a prevalent phenomenon in live streaming.
This contrasts with domains such as movies and books, where users typically consume each item only once, highlighting the unique, ongoing nature of user–streamer interactions in live streaming scenarios.
We further analyze the time intervals between repeat interactions and observe a clear periodic pattern, often occurring on a daily basis.
This suggests that users tend to revisit familiar streamers regularly, possibly due to habit formation or alignment with scheduled broadcast times.
These consistent behavioral patterns underscore the importance of capturing long-term user preferences and temporal rhythms when developing effective recommendation strategies for live streaming platforms.

\begin{table*}[ht]
\centering
\setlength{\tabcolsep}{3pt}
\renewcommand{\arraystretch}{1.2}
\caption{Overall performance of benchmarked models on \ourdata for two recommendation tasks, where the recommended item is either a streamer or a live room. The best and second-best results are highlighted in bold and underlined fonts, respectively. }
\resizebox{.98\textwidth}{!}{
\begin{tabular}{ccccccccccccccc}
\toprule
\multirow{2}{*}{\textbf{Categories}} & \multirow{2}{*}{\textbf{Methods}} & \multicolumn{6}{c}{\textbf{Item: Streamer}} & \multicolumn{6}{c}{\textbf{Item: Room}} \\
\cmidrule(lr){3-8} \cmidrule(lr){9-14} 
 & & Recall@5 & NDCG@5 & Recall@10 & NDCG@10 & Recall@20 & NDCG@20 & Recall@5 & NDCG@5 & Recall@10 & NDCG@10 & Recall@20 & NDCG@20 \\
\midrule
\multirow{4}{*}{\textbf{General}} &\textbf{BPRMF} &0.2929	&0.2176	&0.3841	&0.2470	&0.4850	&0.2724 &\underline{0.2173} &\underline{0.1587}	&\underline{0.2787}	&\underline{0.1785}	&\underline{0.3476}	&\underline{0.1959} \\
&\textbf{NeuMF} &0.2903	&0.2084	&0.3850	&0.2390	&0.4870	&0.2647 & 0.1241	&0.0928	&0.1565	&0.1032	&0.1968	&0.1133\\
&\textbf{LightGCN} &0.2241	&0.1644	&0.2932	&0.1867	&0.3838	&0.2095 &0.2086    & 0.1525  & 0.2722  & 0.1730  & 0.3434 & 0.1910\\
&\textbf{DirectAU} &0.3383	&0.2680	&0.4090	&0.2909	&0.4814	&0.3092 &\textbf{0.2781}	&\textbf{0.2147}	&\textbf{0.3352}	&\textbf{0.2333}	&\textbf{0.3953}	&\textbf{0.2484} \\\midrule
\multirow{4}{*}{\textbf{Sequential}} 
&\textbf{GRU4Rec} &0.3122	&0.2358	&0.3864	&0.2597	&0.4736 &0.2817 &0.1314	& 0.0980	&0.1722	&0.1112	&0.2177	&0.1227\\
&\textbf{SASRec} &0.3773	&0.2966	&0.4582	&0.3228	&\underline{0.5431}	&0.3442 &0.1237	& 0.0933 & 0.1529	& 0.1027	& 0.1943	& 0.1132\\
&\textbf{NARM} &0.3444	&0.2639	&0.4230	&0.2893	&0.5089	&0.3111 & 0.1343	& 0.0980	& 0.1769	& 0.1117	& 0.2226	& 0.1232  \\
&\textbf{Caser} &0.2813	&0.2066	&0.3650	&0.2335	&0.4575	&0.2570 & 0.1247 &0.0894	&0.1641	&0.1020	&0.2116	&0.1140 \\\midrule
\textbf{Time-aware} &\textbf{TiSASRec} &0.3867 &0.3020	&0.4745	&0.3304	&\textbf{0.5607}	&0.3522 & 0.1243	 &0.0943	 &0.1569	 &0.1048	 &0.1964	 &0.1147 \\\midrule
\multirow{2}{*}{\textbf{Generative}} &\textbf{TIGER} &\underline{0.4400}	&\underline{0.3742}	&\underline{0.4830}	&\underline{0.3881}	&0.5246	&\underline{0.3987} & - & - & - & - & - & -\\
&\textbf{LC-Rec} &\textbf{0.4579}	&\textbf{0.3871}	&\textbf{0.4960}	&\textbf{0.3998}	&0.5167	&\textbf{0.4034} & - & - & - & - & - & -\\

\bottomrule
\end{tabular}}
\label{tab:topk}
\end{table*}

\section{Benchmarked Results and Analysis}


In this section, we present the performance of representative baseline methods on top-$K$ recommendation and CTR prediction tasks.

\subsection{Experimental Setups}
\subsubsection{Datasets.}
We conduct our experiments on click behavior, as it provides the most abundant interaction data. 
Following prior work~\cite{rappaz2021recommendation}, recommended items can be defined as either live rooms or streamers, differing in stability: streamer IDs persist across sessions, whereas each live room corresponds to a single session.
As a result, streamer-level interactions are denser, while live room interactions are sparser and more challenging for recommendation models.
We evaluate both item definitions in our experiments.

\subsubsection{Evaluation Protocol.} 
For top-$K$ recommendation evaluation, we adopt a standard leave-one-out splitting strategy.
To simulate real-world settings, candidate items are restricted to those active at the time of each interaction based on their timestamps, from which 10,000 negatives are sampled for each validation and test instance. 
We evaluate models using Recall@\{5, 10, 20\} and NDCG@\{5, 10, 20\}.

For CTR prediction, we combine positive and negative samples and split the data into training, validation, and test sets (8:1:1) chronologically.
User features include genre, age, and follow\_num, while streamer features include genre, age, live\_operation\_tag, and fans\_num.
We evaluate models using AUC and LogLoss.

\subsubsection{Benchmarked Approaches.}
For top-$K$ recommendation, we select representative baselines from four categories: general collaborative filtering methods (BPRMF~\cite{Rendle2009bpr}, NeuMF~\cite{he2017neural}, LightGCN~\cite{he2020lightgcn}, DirectAU~\cite{wang2022towards}), sequential methods (GRU4Rec~\cite{hidasi2015session}, SASRec~\cite{kang2018self}, NARM~\cite{li2017neural}, Caser~\cite{wang2020make}), time-aware methods (TiSASRec~\cite{li2020time}), and generative methods (TIGER~\cite{rajput2023recommender}, LC-Rec~\cite{zheng2024adapting}).

For CTR prediction, we benchmark a set of widely-used models, including FM~\cite{rendle2010factorization}, Wide\&Deep~\cite{cheng2016wide}, DeepFM~\cite{guo2017deepfm}, xDeepFM~\cite{lian2018xdeepfm}, DCN~\cite{wang2017deep}, DCNv2~\cite{wang2021dcn}, DIN~\cite{zhou2018deep}, and DIEN~\cite{zhou2019deep}.

\subsubsection{Implementation Details.}
To ensure fair comparison, we standardize the embedding size to 64 and the batch size to 2048 across all models. 
All methods are optimized using the Adam optimizer, with a grid search performed over the learning rate $\{1e\text{--}3, 1e\text{--}4,5e\text{--}4,1e\text{--}5\}$ and weight decay $\{1e\text{--}5, 1e\text{--}6, 1e\text{--}7\}$, while keeping other hyperparameters as the default settings. 
In sequential recommendation methods, the maximum  history length is set to 50.
Due to computational resource constraints, we only perform streamer prediction for the generative recommendation methods.
All experiments are conducted on NVIDIA Tesla V100 32G GPUs.

\subsection{Experimental Results}
\subsubsection{Top-$K$ Recommendation.}
Table~\ref{tab:topk} presents the overall results of different recommendation methods on \ourdata. 
From the experimental results, we have the following observations and conclusions:

When the recommended items are streamers, LightGCN performs worse than other collaborative filtering-based methods, possibly due to the limited scalability of GNN-based approaches on large-scale datasets.
Moreover, sequential recommendation methods outperform general collaborative filtering-based methods in most cases. 
This may be attributed to the fact that users often watch the same streamer or relevant streamers, resulting in sequential behavioral patterns that sequential models are better equipped to capture. 
Furthermore, time-aware methods such as TiSASRec outperform non-temporal baselines, underscoring the importance of modeling temporal signals in this scenario.
Notably, generative recommendation methods achieve the best overall performance, which may stem from their ability to incorporate semantic information.
This result underscores the importance of leveraging rich auxiliary information rather than relying solely on ID interactions.

When the recommended items are live rooms, we observe a significant drop in performance. 
This can be attributed to the fact that live rooms only exist during the streaming session, and many of them lack interaction history, making them cold-start items. 
Even if a user has previously watched the same streamer multiple times, it is still difficult to generate accurate recommendations based solely on the ID of a newly initiated live room.
Moreover, collaborative filtering-based methods outperform sequential recommendation models in this setting. 
This is because user behavior sequences often do not reflect repeated consumption of the same streamer, making it harder for sequential models to capture user preferences effectively.
These findings highlight the importance of modeling the relationship between live rooms and streamers to improve recommendation accuracy under such dynamic and sparse conditions.

\begin{table}[t]
\caption{Overall performance of benchmarked models on the \ourdata dataset for CTR prediction task. The best and second-best performance methods are highlighted in bold and underlined fonts, respectively.}
	\small
	\centering
        \resizebox{0.9\linewidth}{!}{
	\begin{tabular}{lcccc}
		\toprule
		\multirow{2}{*}{\textbf{Methods}}           & \multicolumn{2}{c}{\textbf{Item: Streamer}} & \multicolumn{2}{c}{\textbf{Item: Room}}\\
        \cmidrule(lr){2-3} \cmidrule(lr){4-5}
        &\textbf{AUC} & \textbf{Logloss} &\textbf{AUC} & \textbf{Logloss} \\
        \midrule
            FM &0.8455	&0.4823 &0.8376 &0.4810\\
            Wide\&Deep  &\textbf{0.8511}	&\underline{0.4011} &0.8385 &\underline{0.4162}\\
            DeepFM &\underline{0.8509}	&\textbf{0.4004} &\textbf{0.8425} &\textbf{0.4130}\\
            xDeepFM &0.8489	&0.4079 &\underline{0.8404} &0.4165\\
            DCN &0.8474	&0.4074 &0.8341 &0.4260\\
            DCNv2 &0.8397  &0.4215 &0.8092 &0.4516\\
            DIN &0.8396	&0.4269 &0.6650 &0.6341\\
            DIEN &0.8362 &0.4297 &0.7480 &0.6176\\
		\bottomrule
	\end{tabular}
	}
	\label{tab:ctr}
\end{table}

\subsubsection{CTR Prediction.}
Table~\ref{tab:ctr} presents the overall results of various CTR prediction methods on \ourdata.
Consistent with the observations in the top-$K$ recommendation task, we find that the performance degrades when the recommended items are live rooms rather than streamers, especially for models like DIN and DIEN that rely heavily on users’ historical behavior sequences.
Moreover, the performance of most CTR models remains relatively similar across the board. 
This may be due to the use of only sparse features, as required by anonymity constraints, which limits the models’ ability to leverage richer semantic information typically carried by dense features.
As a result, methods that rely on the effective fusion of sparse and dense features may fail to realize their full potential.

\section{Potential Research Directions}
Based on the rich user behaviors and multi-granular features provided in \ourdata, we summarize a range of potential research directions that this dataset can support:

\textbf{Top-$K$ Recommendation.}
This is a fundamental task in recommender systems, aiming to generate a ranked list of the top-$K$ most suitable items for each user.
Leveraging the abundant interaction records and precise timestamps provided by \ourdata, researchers can explore a wide range of recommendation techniques, including collaborative filtering \cite{he2017neural,he2020lightgcn,wang2022towards}, sequential recommendation \cite{hidasi2015session,kang2018self,li2017neural,tang2018personalized}, and time-aware recommendation strategies~\cite{li2020time,wang2020make}.

\textbf{XTR Prediction.}
Beyond top-$K$ recommendation, industrial applications often emphasize X-Through-Rate (XTR) prediction tasks~\cite{cao2024moment}, where X refers to specific interaction types, such as Click-Through Rate (CTR) or Gift-Through Rate (GTR). 
CTR prediction focuses on estimating the likelihood of a user clicking a recommended item~\cite{pi2019practice, deng2023contentctr, zheng2023dual}, while GTR prediction estimates the likelihood of a user sending gifts during a live room~\cite{deng2024mmbee,lu2025liveforesighter}.
\ourdata not only provides abundant positive samples for various user behaviors but also includes negative samples, such as unclicked items for CTR and non-gifted items for GTR. 
These features enable a wide range of XTR modeling and evaluation scenarios.

\textbf{Watch Time Prediction.}
In real-world industrial applications, predicting how long a user will stay in a live room is essential for measuring user engagement. 
Watch time is often adopted as a key metric in A/B testing~\cite{cao2024moment,lu2025liveforesighter}, reflecting the effectiveness of recommendation strategies.
This task supports a range of downstream applications such as personalized scheduling, server resource allocation, and streamer performance evaluation.
The detailed records of user watch time provided in \ourdata offer a reliable foundation for developing and evaluating watch time prediction models.

\textbf{Gift Price Prediction.}
Gifting represents a central commercial behavior in live streaming scenarios~\cite{xi2023multimodal,deng2024mmbee}, serving as a major source of income for both streamers and platforms.
Compared to tasks like watch time prediction, gift price prediction poses greater challenges due to the high sparsity and skewed distribution of gift transactions—most gifts are of low value, while high-value gifts occur rarely but contribute disproportionately to revenue.
Accurately estimating user potential gift spending is vital for optimizing monetization strategies and enabling precise user profiling.
\ourdata includes detailed records of gift prices, presenting valuable opportunities to investigate this emerging yet underexplored task in depth.

\textbf{Multi-behavior Modeling.}
Users typically exhibit multiple types of interactions in recommender systems. 
For example, in e-commerce platforms, common behaviors include view, cart, and buy; in live streaming platforms, behaviors extend to click, comment, like, and gift. 
Effectively modeling the relationships among these diverse behaviors has become a research hotspot~\cite{jin2020multi,yuan2022multi,zhang2024saqrec}, as it enables more accurate and context-aware recommendations.
\ourdata provides rich multi-type user behavior data, offering a solid foundation for advancing multi-behavior modeling.

\textbf{Controllable Learning and Recommendation.} 
Real-world recommender systems often need to optimize for multiple objectives simultaneously, such as CTR, GTR, long-view ratio, and total watch time~\cite{liu2025llm,li2025farm}.
Beyond training multi-objective models, a critical challenge lies in test-time controllability: the ability to dynamically rebalance these objectives in response to real-time, context-dependent needs from users or the platform during inference --- e.g., prioritizing GTR amid surges in interactive engagement or boosting long-tail streamer visibility to improve new user retention~\cite{shen2025CL,chen2023controllable}.
The \ourdata dataset offers a rich set of user behaviors and labels, making it well-suited for building and evaluating controllable learning and recommendation architectures in the context of live streaming recommendation.

\textbf{Cold-start Recommendation.}
In real-world applications, new users are constantly joining the platform, and many users may have limited interaction histories.
Cold-start recommendation aims to provide relevant item suggestions for these users with sparse behavioral data, a long-standing and unresolved problem in recommender systems~\cite{lee2019melu,dong2020mamo}.
The \ourdata dataset includes all users and streamers without filtering based on interaction frequency, thereby retaining a substantial proportion of cold-start cases.
This characteristic makes \ourdata a valuable resource for studying and benchmarking cold-start recommendation strategies.

\textbf{Fairness-aware Recommendation.}
To build a more equitable and sustainable live streaming platform, it is essential to consider not only the user experience but also the experience and retention of streamers.
As analyzed in Section~\ref{streamer_analy}, a small number of top streamers receive the majority of user interactions, while many other streamers gain minimal visibility.
Fairness-aware recommendation aims to mitigate systematic biases toward popular streamers and promote the exposure and recommendation opportunities of less popular streamers~\cite{xu2023p,ye2024guaranteeing}.
\ourdata not only provides detailed interaction records for streamers but also includes additional features such as follower counts, which are crucial for developing and evaluating fairness-aware recommendation strategies.


\section{Conclusion}
In this work, we propose a large-scale real-world live streaming dataset, \ourdata, comprising over 23,772 users and 452,621 streamers collected from Kuaishou.
Different from existing datasets, \ourdata records the start and end timestamps of live rooms, captures various user behaviors within the live rooms, and includes rich features for both users and streamers.
These characteristics make \ourdata more representative of real-world live streaming recommendation scenarios.
We further analyze the dataset from both the streamer and user perspectives to reveal the unique characteristics of live streaming scenarios and provide insights that can inform the design of more effective recommendation models.
Furthermore, we evaluate a wide range of representative recommendation and CTR prediction methods on \ourdata, establishing a robust and reproducible benchmark for future research.
Finally, we outline several potential research directions that \ourdata can support. We believe \ourdata will serve as a valuable resource for advancing the study of live streaming recommendation.
\begin{acks}
This work was partially supported by the National Natural Science Foundation of China (No. 62376275, 62472426). 
Work partially done at Beijing Key Laboratory of Research on Large Models and Intelligent Governance, and Engineering Research Center of Next-Generation Intelligent Search and Recommendation, MOE. Supported by fund for building world-class universities (disciplines) of Renmin University of China. Supported by Kuaishou Technology.
\end{acks}

\clearpage
\bibliographystyle{ACM-Reference-Format}
\bibliography{ref}

\end{sloppy}
\end{document}